\documentclass[sigconf]{acmart}
\usepackage{booktabs}
\usepackage[tight,footnotesize]{subfigure}
\usepackage{url}
\usepackage{siunitx}
\usepackage{multirow}

\begin{document}
	\title[H2B]{H2B: Heartbeat-based Secret Key Generation \\ Using Piezo Vibration Sensors}
	\copyrightyear{2019}
\acmYear{2019}
\setcopyright{othergov}
\acmConference[IPSN '19]{The 18th International Conference on Information
Processing in Sensor Networks (co-located with CPS-IoT Week 2019)}{April 16--18,
2019}{Montreal, QC, Canada}
\acmBooktitle{The 18th International Conference on Information Processing in Sensor
Networks (co-located with CPS-IoT Week 2019) (IPSN '19), April 16--18, 2019,
Montreal, QC, Canada}
\acmPrice{15.00}
\acmDOI{10.1145/3302506.3310406}
\acmISBN{978-1-4503-6284-9/19/04}

	\author{Qi Lin$^{12}$, Weitao Xu$^{1\ast}$, Jun Liu$^{1}$, Abdelwahed Khamis$^{12}$, Wen Hu$^{12}$, Mahbub Hassan$^{12}$, Aruna Seneviratne$^{12}$}
	\affiliation{
		\institution{$^{1}$University of New South Wales}
		\institution{$^{2}$Data61 CSIRO}
	}
	\email{(firstname.lastname)@unsw.edu.au}
	
	\renewcommand{\shortauthors}{Q. Lin et al.}

	\begin{abstract}
We present Heartbeats-2-Bits (H2B), which is a system for securely pairing wearable devices by generating a shared secret key from the skin vibrations caused by heartbeat. This work is motivated by potential power saving opportunity arising from the fact that heartbeat intervals can be detected energy-efficiently using inexpensive and power-efficient piezo sensors, which obviates the need to employ complex heartbeat monitors such as Electrocardiogram or Photoplethysmogram. Indeed, our experiments show that piezo sensors can measure heartbeat intervals on many different body locations including chest, wrist, waist, neck and ankle. Unfortunately, we also discover that the heartbeat interval signal captured by piezo vibration sensors has low Signal-to-Noise Ratio (SNR) because they are not designed as precision heartbeat monitors, which becomes the key challenge for H2B. To overcome this problem, we first apply a quantile function-based quantization method to fully extract the useful entropy from the noisy piezo measurements. We then propose a novel Compressive Sensing-based reconciliation method to correct the high bit mismatch rates between the two independently generated keys caused by low SNR. We prototype H2B using off-the-shelf piezo sensors and evaluate its performance on a dataset collected from different body positions of 23 participants. Our results show that H2B has a pairing success rate of $95.6\%$. We also analyze and demonstrate H2B's robustness against three types of attacks. Finally, our power measurements show that H2B is very power-efficient.
\end{abstract}

	

	\maketitle
\vspace{-0.1cm}
\section{Introduction}
\label{sec:introduction}
With recent advances in wireless sensor networks and embedded computing technologies, the popularity of wearable devices, such as FitBit and Apple Watch, is skyrocketing. A recent survey~\cite{seneviratne2017survey} shows that the wearable market is beaming with hundreds of different types of products including smart glasses, smart jewelery, electronic garments, smart shoes, skin patches, and even implanted medical devices (IMDs). We are heading to a future where users are expected to have more than one wearable device continuously monitoring their bodies and providing advanced health and other services~\cite{hiremath2014wearable}. 

In such multi-wearable scenario, wearable devices may deal with private sensory information for a range of tasks, such as data aggregation, sensing coordination, data relaying (from body to cloud). To protect private information, body area networks (BAN) usually rely on symmetric cryptographic keys for secure communication. Traditional key distribution approaches that require pre-deployment in the network initialization stage are not scalable with growing number of wearables. Exploiting biometrics such as gait (e.g, Walkie-Talkie~\cite{xu2016walkie} and BANDANA~\cite{schurmann2017bandana}), Electromyogram (e.g, EMG-KEY~\cite{yang2016secret}), and ECG (e.g., Heart-to-Heart~\cite{rostami2013heart}) as shared key sources has became a hot research field because biometrics is \emph{inherent}, \emph{unique} and \emph{hard to obtain or spoof}.

In this paper, we explore and demonstrate the feasibility of using piezo sensors to detect heartbeat signal. We show that the minor vibrations caused by heartbeats can be detected on multiple body locations such as chest, wrist, waist, neck and ankle. Followed by that, we design and implement a key generation system, H2B, that can exploit the detected heartbeat to generate the same keys for two wearable devices on different body locations of the same user. 

\subsection{Motivations}
\textbf{Why heartbeats?} Although various kinds of biometrics such as gait and Electromyogram (EMG) have been explored to pair devices, there exist some limitations. First, although gait-based authentication/key generation systems have been extensively studied~\cite{xu2016walkie,schurmann2017bandana,xu2017keh}, most existing work validate their systems on a limited dataset. It is still not clear whether gait-based methods apply to large-scale population. In contrast, authors of~\cite{obrist2012cardiovascular,rostami2013heart,cherukuri2003biosec} have shown that inter-pulse interval (IPI), i.e., the interval between two peaks in heartbeat signal, is \emph{highly random} and can be used as a random source to generate keys.

In addition, some systems pose extra burden on the user. For example, gait-based methods require users to walk for a while to generate keys. EMG-based methods require the users to perform a gesture such as clenching the fist as a part of the pairing process. In comparison, heartbeat signal is spontaneous and thus can be measured at \emph{anytime} and in \emph{anywhere}.

\textbf{Why piezo sensors?} In current wearable devices, IPI readings are usually sampled by Electrocardiogram (ECG) or Photoplethysmogram (PPG) sensors. Meanwhile, studies also show that heartbeat signal can be measured by accelerometers in the smartphone and used to extract IPI~\cite{wang2018unlock,ramos2012heart}. Unfortunately, the accelerometer-based methods can \textbf{measure seismocardiogram (SCG) only}, which is chest movement in response to the heartbeat, and fail to detect heartbeats in other locations such as wrist. 
Furthermore, another advantage of piezo vibration sensor over ECG biosensor and accelerometer is its cost-efficiency and energy-efficiency, which makes it more suitable for resource-constrained wearable devices.

Although most current wearable devices are not fitted with piezo sensors, we have observed an increasing number of commercial products equipped with piezoelectric energy harvesters (PEHs) such as AMPY~\cite{AMPY}, SOLEPOWER~\cite{SOLEPOWER}, and KINERGIZER~\cite{KINERGIZER} to reduce dependency on batteries. Recent work has demonstrated that AC voltage generated by PEH can be effectively used for sensing and analyzing body vibrations~\cite{khalifa2018harke}.  Additionally, researchers are considering using wearable piezo sensors to monitor biometrics such as respiration~\cite{jeong2009wearable}. Therefore, we can expect that the PEH or piezo sensors will be embedded in wearable devices to enable more applications in the future.

An inherent problem with biometrics is that the measurement is never perfect~\cite{cherukuri2003biosec}. Parallel measurements of the same source by multiple sensors at the same time differ from each other, though they maintain a certain proximity. Unlike ECG sensors or accelerometers, piezo sensors are not inherently designed for precise motion detection. Therefore, the IPI measured by the piezo sensors contains significant noise which becomes the key challenge for designing a system that requires 100\% bit match between the keys generated by two devices independently. The low SNR poses two challenges. Firstly, as the measurements contain significant random noise, the amount of entropy in them,
which is critically important for the quality of the keys, will be reduced significantly. Secondly, the noisy measurements lead to significant number of bit mismatches between the two keys. The mismatches will either cause key generation failure or increase the time required to generate matching keys, which in turn decreases user experience.

To this end, we apply a quantile function-based quantization method to fully extract useful entropy, and propose a novel Compressive Sensing (CS)-based reconciliation method to correct mismatches in quantized bits. To the best of our knowledge, this is the first work that explores and demonstrates the feasibility of using piezo vibration sensors to detect heartbeat signal and use the signal to generate keys.

The contributions of this work can be summarized as follows:
\begin{itemize}
	\item We conduct the first study to experimentally analyze the potential of piezo-based IPIs. Our result shows that piezo-based IPIs have smaller entropy and higher mismatch rate than those measured by ECG. As a result, the shared key generation using the piezo-based IPIs takes longer time and has a lower success rate. [Section~\ref{sec:key_potential}]
	\item We propose H2B, a symmetric key generation system exploiting IPIs captured by polyvinylidene difluoride (PVDF) piezo sensor. We apply a quantile function-based quantization to fully extract the entropy and propose a novel \emph{CS-based reconciliation} to overcome the high level of variation inherent in the piezo sensor's measurements.[Section \ref{sec:ptcl}]
	\item We provide a proof of concept implementation of H2B using off-the-shelf piezo sensors and evaluate H2B with 23 participants. Our results show that: 1. Quantile function-based quantization can extract $2.9$ bits entropy from each IPI while a naive Gray-coded representation can extract $1$ bit only; 2. CS-based reconciliation significantly improves the $128$-bit key generation success rate from a low $34.2\%$ to a $95.6\%$ compared to traditional error correction methods. [Section \ref{sec:hardware}]
	\item We conduct comprehensive attack analysis to show that H2B is highly robust against typical attacks.[Section \ref{sec:ev_security}]
	\item Finally, we demonstrate that H2B is very power-efficient for sampling IPI and generating symmetric keys.[Section \ref{sec:power}]
\end{itemize}

\section{Feasibility Study}
\label{sec:key_potential}
\subsection{ECG}
\label{sec:bg_ipi}
\begin{figure}[h!]
	\centering
	\subfigure[Typical ECG waveform and IPI]{ 
		\label{fig:ecg}
		\includegraphics[width=7cm]{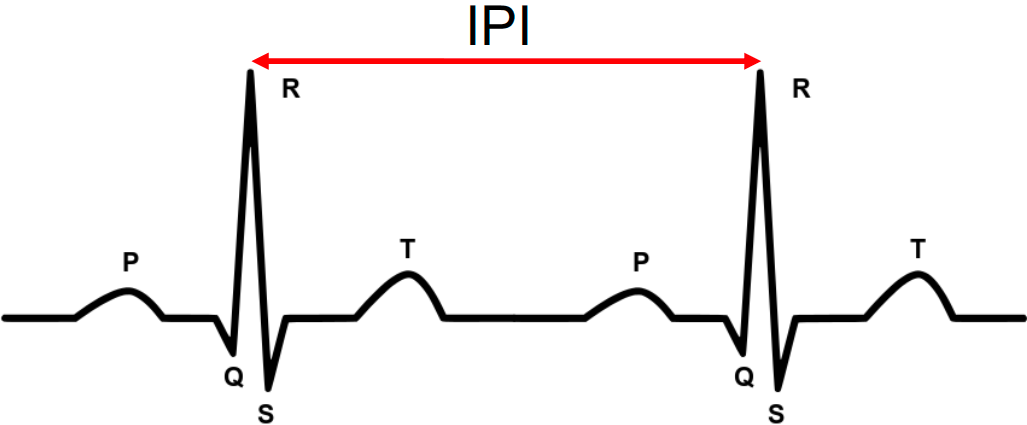}}
	\subfigure[An IPI measured from chest (captured by a piezo sensor) ]{ 
		\label{fig:ipi-chest} 
		\includegraphics[width=8cm]{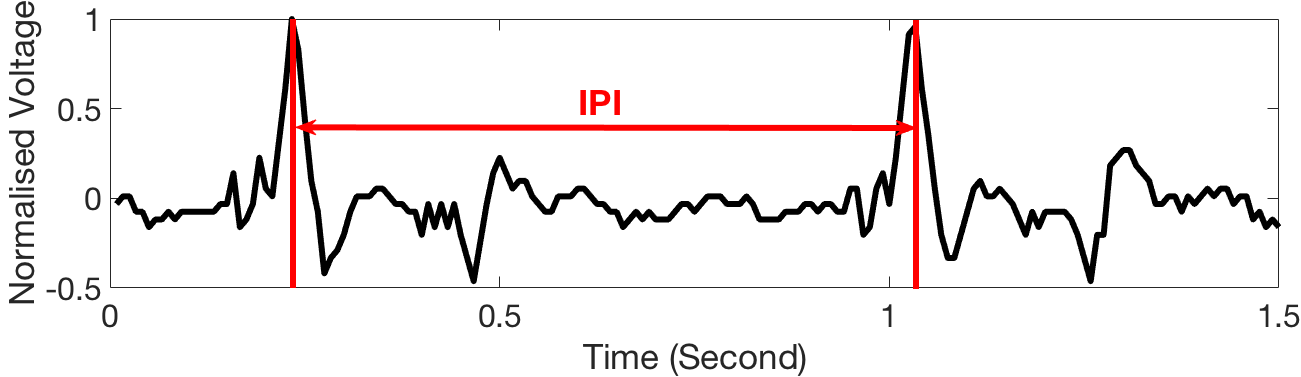}}
	\subfigure[An IPI measured from wrist (captured by a piezo sensor)]{ 
		\label{fig:ipi-wrist} 
		\includegraphics[width=8cm]{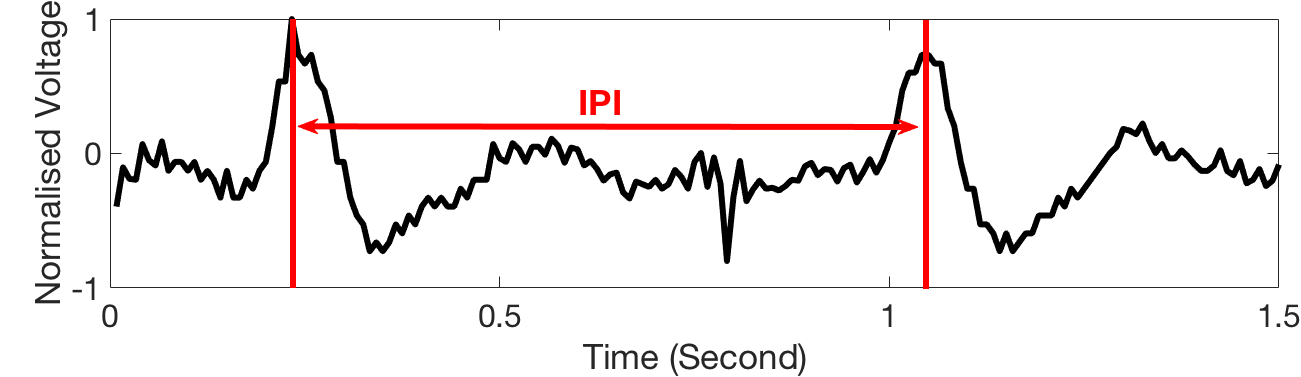}}
	\caption{IPIs in different representations.}
    \label{fig:IPI}
    	    \vspace{-0.2in}
\end{figure}

Figure~\ref{fig:ecg} depicts two typical ECG waveforms of a
healthy person. A heartbeat cycle in ECG is characterized by the combination of three graphical deflections (Q, R and S waves), which is also known as QRS complex. The R-peak is the most crucial feature of an ECG waveform as it corresponds to the ''beat'' in a heartbeat. The time interval between two consecutive R-peaks is referred to inter-pulse interval (IPI).

The principle of using piezo vibration sensors to detect IPIs can be explained as follows. On the chest and waist, the heart contraction will cause regular vibrations in the chest which is called seismocardiogram (SCG). Such vibrations can be detected by the piezo vibration sensor attached to the chest and waist as shown in Figure~\ref{fig:ipi-chest}. After each heartbeat, the blood is ejected from the heart into arteries all over the body. In order to accommodate the blood, the arteries need to expand and recoil regularly which produce ballistocardiogram (BCG) signal~\cite{dart2001pulse}. The expansion and recoiling effect, also known as the blood pressure pulse, can be measured on skin near arteries (e.g. the radial artery in the wrist). Therefore, if we attach a piezo sensor near the artery in other locations (e.g., wrist), we can extract IPI information from the measurements as shown in Figure~\ref{fig:ipi-wrist}.

\begin{figure}[h!]
	\centering
	\includegraphics[width = 8cm]{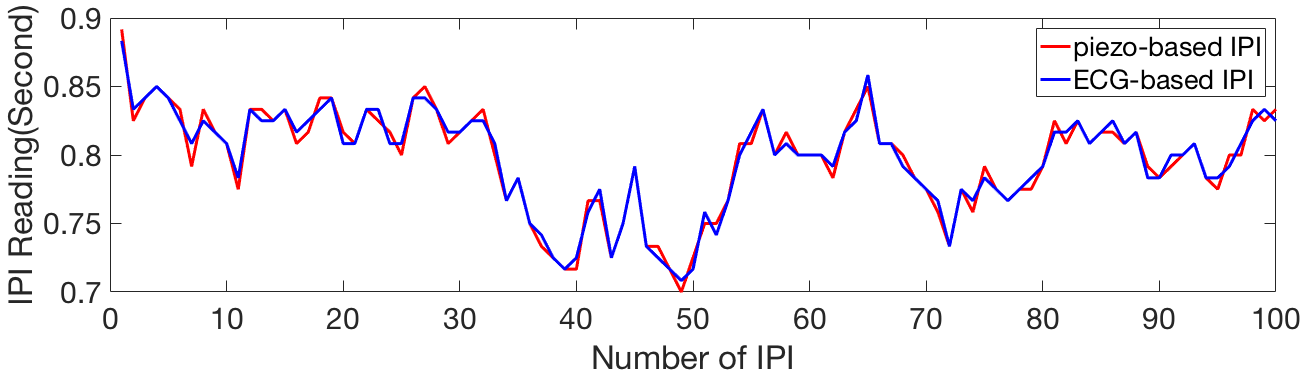}
	\caption{Piezo sensor v.s. ECG biosensor.}
	\label{fig:ecgvspiezo}
		    \vspace{-0.2in}
\end{figure}

From Figure~\ref{fig:IPI}, we also find that the curves of heartbeat cycle measured from different body locations do not follow the same pattern as typical QRS complex. Nevertheless, we can apply a local peak point (either maximum or minimum) to represent each heartbeat and further extract IPIs. Then there is a question: \emph{How accurate is the IPI measured by a piezo sensor?} To this end, we compare the measurement precision of a piezo sensor with a commercial heart rate monitor device Polar H10\footnote{\url{https://www.polar.com/au-en/products/accessories/h10_heart_rate_sensor}} as the ground-truth. Our results from 23 participants show that the RMSE of IPIs measured by piezo sensors is 6ms only (dataset is explained in Section~\ref{sec:exp}). Figure~\ref{fig:ecgvspiezo} plots a series of IPI measured by the piezo sensor and the ECG biosensor (i.e., Polar H10) respectively. We can see that piezo-based IPIs are very close to those measured by the ECG biosensor. 

Based on the analysis above, it is feasible to use piezo sensors to measure accurate IPIs. However, there still exist two problems. Firstly, the IPIs measured form different body locations are slightly different. For example, the IPI measured in Figure~\ref{fig:ipi-wrist} (wrist) is slightly longer than that in Figure~\ref{fig:ipi-chest} (chest). Secondly, the IPIs measured by the piezo sensors has some small deviations as can be seen from Figure~\ref{fig:ecgvspiezo}. These two problems will lead to the low entropy in the signals and high mismatch rate 
between the two keys, which we will discuss next.

\subsection{Piezo-based IPI as Secret Source}
\label{sec:ipi_benchmark}
The secret source for symmetric key based secure communication protocols must satisfy \textbf{randomness} and \textbf{proximity} conditions~\cite{cherukuri2003biosec}. Firstly, for security purpose, the generated key must be cryptographically random, or independently and identically distributed (i.i.d.). Secondly, for usability purpose, the final keys generated from the same source independently by two legitimate parties (i.e., Alice and Bob) must agree. In terms of randomness, IPI is random according to the features of heart rate variability~\cite{oweis2014qrs}. Furthermore, previous studies~\cite{xu2011imdguard,rostami2013heart} on PhysioBank\footnote{https://physionet.org/physiobank/} database have demonstrated that IPI can be used to extract random keys. In terms of proximity, IPI measured from different locations are close to each other because, intuitively, they measure the same heartbeat signal. Therefore, IPI measured by piezo sensors at least contains a certain amount of information that satisfies both randomness and proximity conditions. 

To this end, we benchmark piezo-based IPIs against ECG-based IPIs to show the potential of piezo-based IPIs as a secret source. We built several prototype wearable devices and collected a preliminary dataset. 
We use Heart2Heart (H2H)~\cite{rostami2013heart}, which is a state-of-the-art ECG-based system, as a baseline and follow its quantization process, which uses the Gray-code representation of raw IPI values. 

We use entropy and mismatch rate as the evaluation metrics. The entropy denotes information contains in each bit. To satisfy the randomness condition, the entropy for each extracted bit must be sufficiently high. We empirically select an entropy threshold of 0.95 to determine valid bits. The mismatch rate is the probability of mismatched bits over all quantization bits produced from the IPI measurements captured by Alice and Bob respectively. To satisfy the proximity condition, the mismatch rate must be sufficiently small. We empirically select an mismatch rate threshold of $20\%$ as the proximity condition. Table~\ref{tab:bit_comp} shows the entropy and the mismatch rate of Gray-coded bits of piezo-based and ECG-based IPI respectively. In terms of the entropy, both piezo-based IPI and ECG-based IPI have 5 high-entropy bits (Bit 1-5) that can be used as valid keys by satisfying the randomness condition of $\geq 95\%$ entropy per bit. In terms of the mismatch rate, however, bits generated from the piezo-based IPI have much higher mismatch rates than those of the ECG-based IPI. For example, the mismatch rate of Bit 4 is $32.8\%$ for the piezo-based IPI while the same bit extracted its ECG-based counterpart has $0.9\%$ mismatch rate only. 
Therefore, if we adopt the method in~\cite{rostami2013heart} directly, we have to discard 4 least significant bits and keep Bit 5 only to satisfy the proximity condition with an mismatch rate of $\leq 20\%$. The result highlights that piezo-based IPIs as a secret source has a very significantly lower SNR compared to ECG, thus faces the challenges of \textbf{significantly lower entropy and higher bit mismatch rate}.

\begin{table}[h!]
	\centering
	\caption{Comparison of entropy and mismatch rate.}
	\label{tab:bit_comp}
	\resizebox{8.5cm}{!}{
		\begin{tabular}{ccccc}
			\toprule
			\multirow{2}{*}{Bit}  & \multicolumn{2}{c}{piezo-based IPI}  & \multicolumn{2}{c}{ECG-based IPI (H2H Results)}  \\
			\cline{2-5}
			\multirow{2}{*}{} & Entropy & Mismatch Rate & Entropy & Mismatch Rate \\ 
			\hline
			8 (MSB) & 0.05 & 0$\%$ & 0.27 & 0.1$\%$ \\
			7 & 0.44 & 1$\%$ & 0.80 & 0.3$\%$ \\
			6 & 0.64 & 6.9$\%$ & 0.90 & 0.4$\%$ \\
			5 & 0.98 & 14.4$\%$ & 0.98 & 0.6$\%$ \\
			4 & 1 & 32.8$\%$ & 1 & 0.9$\%$ \\
			3 & 1 & 42.8$\%$ & 1 & 1.8$\%$ \\
			2 & 1 & 45.2$\%$ & 1 & 3.9$\%$ \\
			1 (LSB) & 1 & 49.8$\%$ &  1 & 8$\%$ \\
			\bottomrule
	\end{tabular}}
	    \vspace{-0.1in}
\end{table}

\section{System Model}
\label{sec:models}
\subsection{Trust Model}
\label{sec:op_models}
\begin{figure}
	\centering
	\includegraphics[width = 6cm]{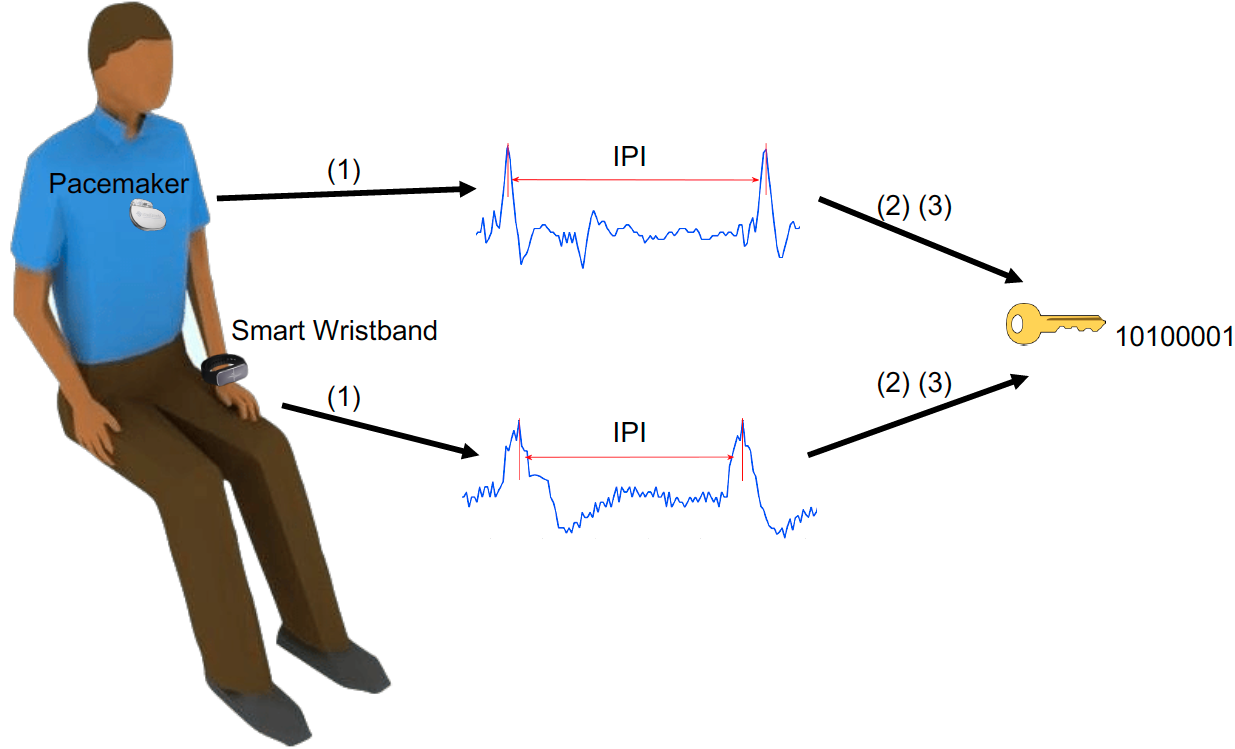}
	\caption{System Overview. (1) IPI extraction (2) Quantization (3) Reconciliation.}
	\label{fig:model}
		    \vspace{-0.2in}
\end{figure}

We envision the usage of H2B primarily for wearable devices that have close contact with user's body as these devices can detect and measure user's heartbeat pulses and further use them to establish common cryptographic keys for secure communications. For example, as shown in Figure~\ref{fig:model}, a user (Joe) gets up from bed and wears a smart wristband (Alice) on his wrist. He wishes to read some information from his pacemaker (Bob) which is inside his chest. Therefore, Joe launches H2B APP in the wristband, and after a short while Alice and Bob automatically establish a secret key by measuring Joe's IPI. Finally, Bob transfers private data to Alice via the established secure channel protected by the common secret key. In a mobile environment, the demand for this type of temporal association is increasing. Therefore, H2B is well suited for frequent and short-lived data exchange between wearable devices. Moreover, it provides an unobtrusive, friendly, spontaneous key generation approach for users. 

Same as previous biometrics-based key generation systems~\cite{xu2016walkie,rostami2013heart}, we assume that all on-body devices that can measure the heartbeat signals of a user are trustworthy. As discussed in~\cite{mayrhofer2009shake}, an attacker can neither control the vibration recorded
locally by these devices nor perfectly estimate it; otherwise
the protection of legitimate devices is impossible. The user is assumed to be capable of detecting such attacks if an attacker tries to modify the on-body devices or place an adversary device on her. In fact, there are a number of approaches to protect such attacks. For example, in order to make sure the devices have not been modified, the device manufacturers may apply tamper-resistant techniques~\cite{ravi2004tamper}. ARM TrustZone extension may also be used to ensure the integrity of the sensors~\cite{guan2016physical,liu2012software}.

\subsection{Adversary Model}
\label{sec:model_attack} 
We assume the presence of a strong attacker Eve during the key generation process of H2B. Eve has full knowledge of the system and control of the communication channel, i.e., she may monitor, jam, and modify messages at will. The detection of jamming attacks is out of the scope of  this paper as jamming attacks have been extensively studied and we may adopt the methods in~\cite{osanaiye2018statistical,xu2005feasibility}. Compromising integrity such as modifying messages can be prevented by implementing Message authentication code (MAC) in transmitted messages. Apart from these attacks, there are three types of attacks particularly targeting H2B that may be used by Eve to recover the secret key: passive eavesdropping attacks, active presentation attacks and active video magnification attacks. 

\begin{enumerate}
	\item \textbf{Passive eavesdropping attacks.} A passive attacker understands the key generation mechanism and may monitor the communication between Alice and Bob. By eavesdropping the public components of key generation, Eve tries to generate a key based on her own heartbeat information and use this key to pair with one or both of the legitimate devices. In order to counter this type of attacks, H2B needs to make sure the extracted keys that are \textbf{distinctive} among different users.
	\item \textbf{Active presentation attacks.} A presentation attacker may access historical IPI data of a user. Eve attempts to pair with legitimate devices by submitting key generated from historical data. In order to counter this type of attacks, the keys generated by H2B need to provide \textbf{forward secrecy}.
	\item \textbf{Active video attacks.} A recent work presented a video-based head pulse motion detection method~\cite{balakrishnan2013detecting}, which is able to detect heartbeat signals by magnifying subtle head motion caused by the influx of blood at each heartbeat. Therefore, an active video attacker may observe and film the head of a user, and use this method to extract IPIs to generate the same key as that produced by a legitimate device. Active video attacks may be considered as one form of \textbf{side-channel attacks}.
\end{enumerate}
We will provide a more detailed discussion on the robustness of H2B against these attacks in Section~\ref{sec:ev_security}.

\section{Protocol Design}
\label{sec:ptcl}
As shown in Figure~\ref{fig:model}, the protocol comprises of three stages: IPI extraction, quantization, and reconciliation. In the IPI extraction stage, we collect heart beat signal via  piezo sensors and extract a series of consecutive IPIs. In quantization stage, the IPIs are quantized into bit vectors. IPI extraction and quantization stages are executed by Alice and Bob independently. Due to the measurement noise, there are usually mismatches in the quantized bits between the bit vectors produced by Alice and Bob respectively. Therefore, in the reconciliation stage Alice and Bob exchange a certain amount of information to correct all mismatches and generate a bit-for-bit matching key. 
Before we discuss the three stages of H2B in details, we summarize the mathematical symbol notations used in this paper 
in Table \ref{tab:notation}.
\begin{table}[h!]
	\centering
	\caption{Symbol notations}
	\label{tab:notation}
	\resizebox{8.5cm}{!}{
		\begin{tabular}{|c|l|}
			\hline
			Symbol  & Meaning\\
			\hline
			\hline
            $F(\cdot)$ & cumulative distribution function \\
            $I(\cdot)$ & quantile function or inverse cumulative distribution function \\
			$\chi$, $X$ & original IPI value and its domain (e.g. time)\\
			$T_{i}$ & quantization threshold \\
			$n$ & number of quantized IPI values \\
			$H(\cdot)$ & entropy function \\
			$N$  &  signal length of bit vector\\
			$M$ & length of compressed key\\
            $Q$ & security threshold \\
            $P$ & effective threshold \\
			$x$ &  bit vector (or raw key) after quantization\\
			$y$  & compressed key\\
			$\Phi$ & compressing matrix \\
			$S$ & sparsity, the number of non-zero elements in a vector\\
			$S_{Alice}$ & sparsity of key Alice $x_{Alice}$\\
			$S_{\Delta A, B}, S_{\Delta A, E}$ & sparsity of mismatch vector between Alice and Bob, and between Alice and Eve \\
			\hline
	\end{tabular}}
	    \vspace{-0.2in}
\end{table}

\subsection{IPI Extraction}
As discussed earlier in Section~\ref{sec:bg_ipi}, although piezo sensors at different locations can capture heartbeat signal simultaneously, the measurements do not perfectly match and contain significant noise. To accurately extract IPI values, we first apply a Savitzky-Golay (SG) filter on the raw signal to reduce the impact of noise, which can smooth the original curves without changing the peak locations in the curves. Then, we locate each heartbeat using a local peak detection algorithm. IPI values can be finally computed from the time difference of consecutive heartbeats.

Synchronization is required to ensure all legitimate devices capture IPIs at the same time. In H2B, the synchronization of Alice and Bob is achieved using a time-slotted channel hopping-based (TSCH)~\cite{ieee2011ieee} time synchronization mechanism\footnote{Other time synchronization methods such as Cheepsync~\cite{Cheepsync} may also be used.}. With TSCH, the timers in the devices are synchronized when they join a network. 

\subsection{Quantization}
A common method of converting IPIs into bits is directly using the Gray-code representation of IPI values~\cite{rostami2013heart,bao2008using}. However, as discussed in~\cite{xu2011imdguard}, the keys generated from such method are actually not random because IPI values fluctuates around the average value and follow a normal distribution. Therefore, attackers may perform statistical analysis and gain significant amount of knowledge about the key. For example, since the average of IPIs is 850ms, let us assume the first IPI value is 840ms and thus its corresponding Gray code is ''\underline{10111}01100''. If the second IPI value is 860ms, its corresponding Gray code is ''\underline{10111}10010''. We can see the the first 5 digits of the two produced Gray codes are exactly the same. Therefore, the randomness lies only in the lower digits. To address this problem and fully exploit the potential of
keying material, we first convert the IPI readings into a dataset that follows a uniform distribution. Then the converted IPI values are quantized into bits using Gray code representation. Since the data used to generate the final key follow a uniform distribution instead of a normal distribution, all bits in the extracted keys have high entropy. 

In details, we start at converting the raw IPI readings which follow a normal distribution into a dataset that follows uniform distribution. We use a quantile function approach similar to~\cite{xu2011imdguard}. Specifically, we segment the normal distribution into a set of $n$ discrete values with an equal probability $\frac{1}{n}$. Suppose we measure a series of IPI readings $X=\{\chi_{1},\chi_{2},\cdots,\chi_{N}\}$. The cumulative distribution function (CDF) of these readings is $F(\chi)$. The function $F(\chi)$ gives us the probability $p$ that $\chi$ will take a value $\leq \chi$. Instead, the quantile function which is defined as $I(p)=inf\{\chi\in \mathbb{R}: p\leq F(\chi)\}$ does the opposite: it produces a threshold value $\chi$ below which the random draws from $\chi$ will fall into $p$ percent of the time. Therefore, we can simply obtain the thresholds by choosing $p=(\frac{1}{n}, \frac{2}{n}, ..., 1)$ as follows:
\begin{equation}
T_{i}= I(p) \qquad p=(\frac{1}{n}, \frac{2}{n}, ..., 1)
\label{eq:normal}
\end{equation}
where $T_{i}$ is the $i$-th threshold. The design of quantization now becomes finding a proper $n$. As the extracted keys cannot exceed the theoretic entropy in the source, the number of quantized IPI values is thereby determined by the total entropy of the original IPI values. Therefore, we choose $n = 2^{\lfloor H(X)\rfloor}$ where $H(X)$ is the entropy of IPI values and $\lfloor H(X)\rfloor$ is the maximum integer smaller than $H(x)$.

\begin{figure}[h!]
	\centering
	\subfigure[Histogram of original IPI]{ 
		\label{fig:h1} 
		\begin{minipage}[b]{0.25\textwidth} 
			\centering 
			\includegraphics[width=4cm]{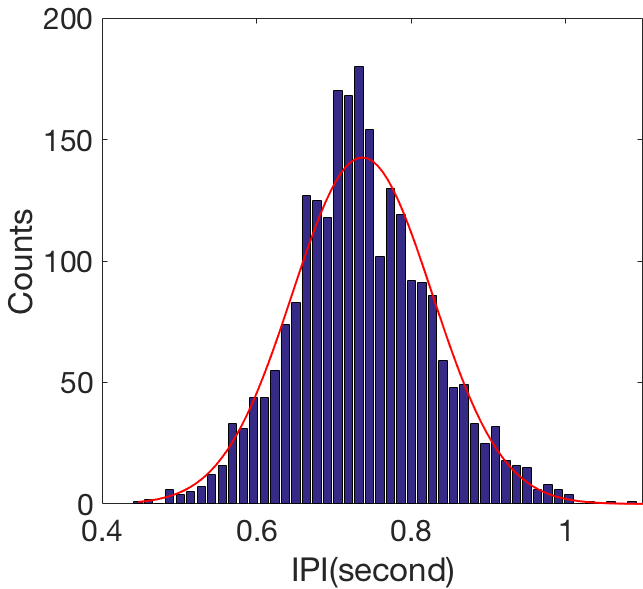}
	\end{minipage}}%
	\subfigure[Histogram of quantized IPI]{ 
		\label{fig:h2} 
		\begin{minipage}[b]{0.25\textwidth} 
			\centering 
			\includegraphics[width=4cm]{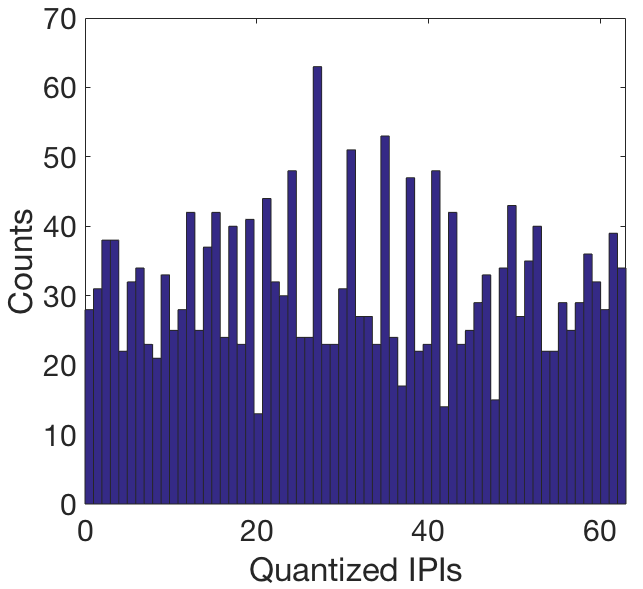} 
	\end{minipage}}%
	\caption{Illustration of quantization: original IPIs follow a normal distribution. After quantization, the quantized IPIs roughly follow a uniform distribution.}
	\label{fig:hist}
	    \vspace{-0.1in}
\end{figure}

After finding the thresholds, the original IPI values falling into the $i$-th segmentation are quantized into $i-1$. For example, the IPI values falling into the first segmentation are quantized into 0, and the IPI values falling into the last segmentation are quantized into $n-1$. In order to distinguish IPI values before and after quantization, we term the IPI values after quantization as quantized IPIs. Figure~\ref{fig:hist} illustrates this process. Figure~\ref{fig:h1} shows that the original IPI values follow a normal distribution similar as~\cite{xu2011imdguard}. After quantization, the quantized IPIs roughly follow a uniform distribution as shown in Figure~\ref{fig:h2}. The mismatches are caused by the noise.

Finally, we use Gray code to encode the quantized IPI values. After this step, each IPI value is represented by a bit strings containing `0' and `1'. Then, all the bit strings generated from collected IPIs are concatenated to form an initial key $x_{Alice}$ and $x_{Bob}$, which are generated on two devices independently. 
 
Here we briefly illustrate the advantage of this quantization process. We apply the above quantization process on the same data used in Section~~\ref{sec:ipi_benchmark} and plot the results in Table~\ref{tab:quan_bit}. Compared to the results in Table~\ref{tab:bit_comp}, though our method can generate 6 bits per IPI only, we find that the first 3 most significant bits (i.e.m Bits 4, 5, 6) have high entropy and low mismatch rates. The result suggests that our method improves key generation rate by 3 times because a naive Gray code representation can only generate 1 bit with high entropy (Bit 5) based on the analysis in Section~~\ref{sec:ipi_benchmark}. However, quantization does not remove the random noise of original IPI values completely. From Table~\ref{tab:quan_bit}, we can see random noise still cluster in least significant bits (Bit 1-3) after quantization and causes high mismatch rates. Therefore, H2B discards the least 3 bits and keep Bits 4 to 6 only. 
In other words, H2B extracts 3 bits per IPI. In next section, we will address the challenge of the mismatches (between 10\% and 20\%) in Bits 4 to 6. 

\begin{table}[h!]
	\centering
	\caption{Results of quantization process.}
	\label{tab:quan_bit}
	\resizebox{5.5cm}{!}{
		\begin{tabular}{ccc}
			\toprule
			{Bit}   & Entropy & Mismatch Rate\\ 
			\hline
			6 (MSB) & 0.96 & 12.9$\%$ \\
			5 & 0.96 & 12.4$\%$ \\
			4 & 1 & 19.4$\%$ \\
			3 & 1 & 35.8$\%$ \\
			2 & 1 & 34.8$\%$ \\
			1 (LSB) & 1 & 43.2$\%$ \\
			\bottomrule
	\end{tabular}}
\end{table}
\vspace{-0.2in}
\subsection{Reconciliation}
\label{sec:rec}
\subsubsection{Principles of Compressive Sensing}
\label{subsec:cs}
Before we introduce the CS-based reconciliation method to correct mismatches, we first review the principles of CS briefly. 
CS is an information theory~\cite{wang2010information,baraniuk2007compressive} that proposes
a method to recover a high dimension \textbf{sparse} signal from a small number of (i.e., low dimension) measurements. Let $x \in \mathbb{R}^{N}$ be an unknown data vector, $y \in \mathbb{R}^{M}$ be a measurement vector, and $\Phi \in \mathbb{R}^{M \times N}$ be a projection matrix from a higher dimension ($N$) to a lower dimension ($M$), where $M << N$. Here, both vector $y$ and matrix $\Phi$ are given, and the unknown vector $x$ needs to be determined. We can then write the problem in a linear form as:
\begin{equation}
y = \Phi x.
\label{eq:y}
\end{equation}
Since $M < N$, Eq. (\ref{eq:y}) is under-determined, and is impossible to be solved in a general form. 
\begin{figure*}[th]
	\centering
	\includegraphics[width = 14cm]{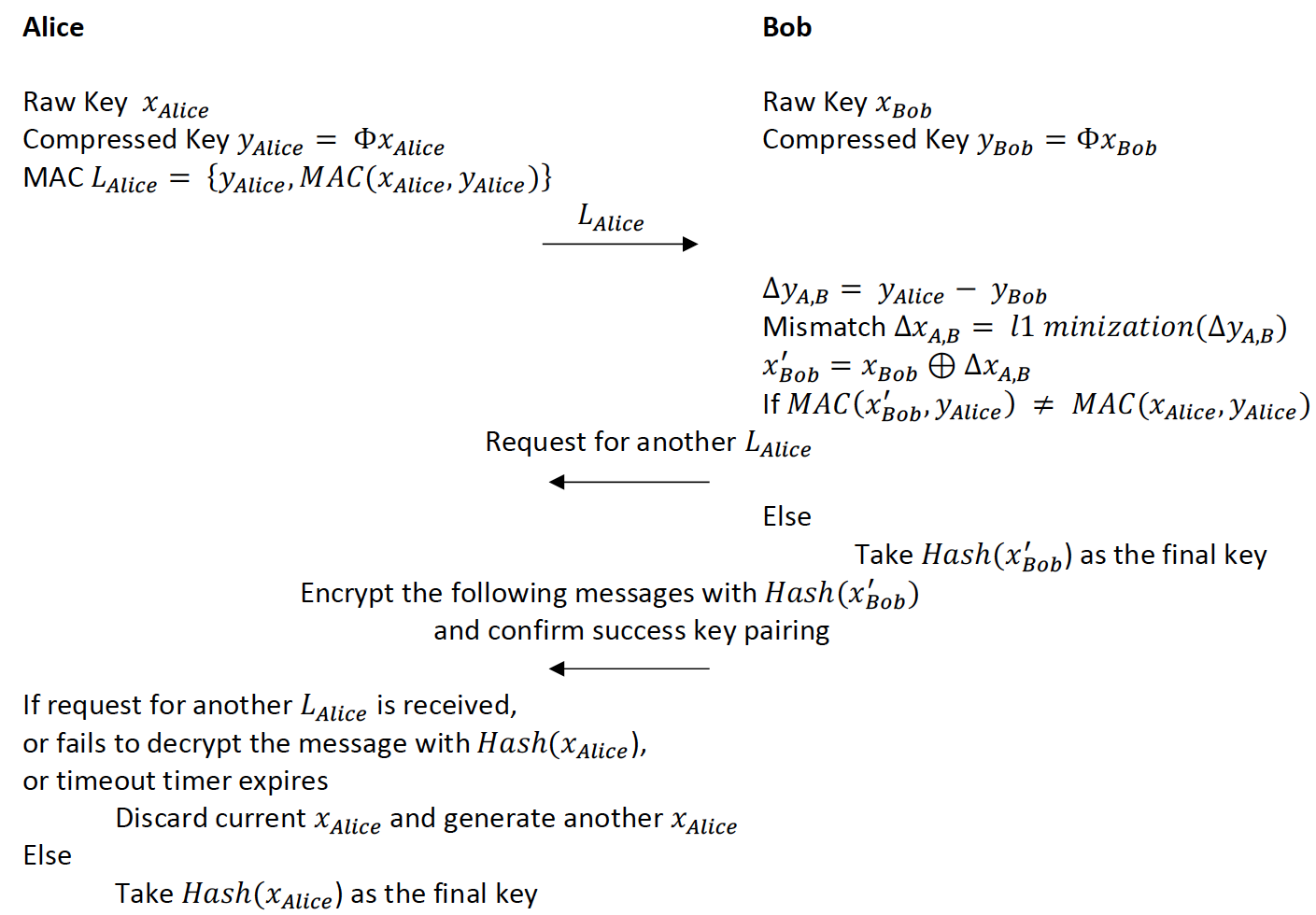}
	\caption{Reconciliation protocol.}
	\label{fig:ptcl_rec}
		    \vspace{-0.2in}
\end{figure*}

CS imposes the requirement that data vector $x$ is \emph{sparse}; namely, most of the elements in $x$ are zeros. Let
$S$ denote the number of non-zeros in $x$, then $x$ is sparse if $S << N$. $S$ in CS is termed as \textbf{sparsity}. CS theory states that we can recover data vector $x$ with an overwhelming probability by solving the following
$\ell_1$ minimization problem:
\begin{equation}
\hat{x} = \arg\min\limits_{x} \lVert x \rVert_{1} \text{ subject to } \lVert y  - \Phi x \rVert_2< \epsilon,
\label{eq:l1:original}
\end{equation}
where $\epsilon$ is noise and provided \textbf{Conditions C1 and C2} as follows.
\begin{itemize}
	\item \textbf{C1}. $\Phi$ must obey the restricted isometry property (RIP). For example, $\Phi$ satisfies
	RIP if each element in $\Phi$ is $\pm1$ with equal probability, i.e., symmetric Bernoulli distribution. 
	\item \textbf{C2}. $M  > S * log(N/S)$. 
\end{itemize}    
Note that \textbf{C2} is a \emph{sufficient condition}, and Wang et. al show that the \emph{necessary condition} is~\cite{wang2010information}:
\begin{itemize}
	\item \textbf{C3}.  $M > S$. 
\end{itemize}

A symmetric key secure communication protocol requires a bit-for-bit equal key for Alice and Bob. After quantization, Alice and Bob generate the bit vectors ($x_{Alice}, x_{Bob} \in  \mathbb{Z}_{2}^{N}$) respectively. However, we usually have $x_{Alice}\approx~x_{Bob}$ in practice (see Table~\ref{tab:quan_bit}). In the context of information reconciliation, our intuition is that \textbf{the mismatched vector between two keys $x_{Alice}$ and $x_{Bob}$ is sparse as they are very close to each other (e.g., $\leq 20\%$ for Bits 4 to 6 in Table~\ref{tab:quan_bit}), therefore can be recovered from Eq. (\ref{eq:l1:original}) efficiently}.

\subsubsection{Information Reconciliation Protocol with Compressive Sensing}
\label{subsec:rec}
Figure~\ref{fig:ptcl_rec} illustrates the flowchart of proposed information reconciliation protocol. The roles of Alice and Bob are interchangeable. The proposed method is inspired by the application of CS in error correction~\cite{fu2013compressed}, but our approach is based on different assumption (i.e., the mismatched vector is sparse) and is therefore different. In fact, the proposed method may be applied in all kinds of symmetric key generation system for information reconciliation.

Firstly, Alice and Bob derive their own key candidate vectors $x_{Alice}$ and $x_{Bob}$ from local observation independently. Secondly, they \emph{project} the vectors to lower dimension spaces  (i.e., from $N$ to $M$,  where $N >> M$), with Eq. (\ref{eq:y}) respectively, namely $y_{Alice} = \Phi x_{Alice}$ and $y_{Bob} = \Phi x_{Bob}$, where $\Phi$ is a known key generation protocol parameter (i.e., a known Bernoulli matrix consisting of $\pm1$ with equal probability). Thirdly, Alice transmits $y_{Alice}$ to Bob via an unauthenticated channel. On receiving $y_{Alice}$, Bob calculates the mismatched vector $\Delta y_{A, B} = y_{Alice} - y_{Bob}$ in a lower dimension space. According to the proximity condition in Section~\ref{sec:ipi_benchmark}, the mismatched vector in a higher dimension space $\Delta x_{A, B} = x_{Alice} \oplus x_{Bob}$ will have less than $20\%$ non-zero elements only (\textbf{sparse!}), i.e., a small $S_{\Delta A, B}$. Furthermore, if project matrix $\Phi$ satisfies Conditions \textbf{C1} and \textbf{C2} in Section~\ref{subsec:cs}, $\Delta x_{A, B}$ can be recovered using Eq. (\ref{eq:l1:original}) as:
\begin{multline}
	\hat{\Delta x_{A, B}} = \arg\min\limits_{\Delta x_{A, B}} \lVert \Delta x_{A, B}\rVert_{1} \quad s.t.
	\lVert \Delta y_{A, B} \oplus  \Phi \Delta x_{A, B} \rVert_2 < \epsilon. 
	\label{eq:l1}
\end{multline}
Finally, with $\Delta x_{A,B}$, Bob can correct mismatched bits by deriving a new key $x^\prime_{Bob} = x_{Bob}\oplus\Delta x_{A, B}$, and $x^\prime_{Bob} = x_{Alice}$.  

As the compressed key $y_{Alice}$ is transmitted via a public channel, Eve may eavesdrop it. Therefore, there are two obvious potential vulnerabilities of this reconciliation protocol.

\textbf{Vulnerability 1}: Eve may reconstruct $x_{Alice}$ from $y_{Alice}$ directly using Eq.(\ref{eq:l1:original}). 

\textbf{Vulnerability 2}: Eve may perform the three types of attacks discussed in Section~\ref{sec:model_attack} to obtain IPI values. Then she can use them to derive her own key, i.e., $y_{Eve}$, to calculate $\Delta y_{A, E} = y_{Alice} - y_{Eve}$, where $y_{Eve} = \Phi x_{Eve}$. Therefore, she may recover $\Delta x_{A, E}$ with Eq. (\ref{eq:l1}) by replacing $\Delta x_{A, B}$ and $y_{A, B}$ with $\Delta x_{A, E}$ and $y_{A, E}$ respectively, and obtain $x_{Alice}$. 

\begin{figure}[ht]
	\centering
	\includegraphics[width = 5cm]{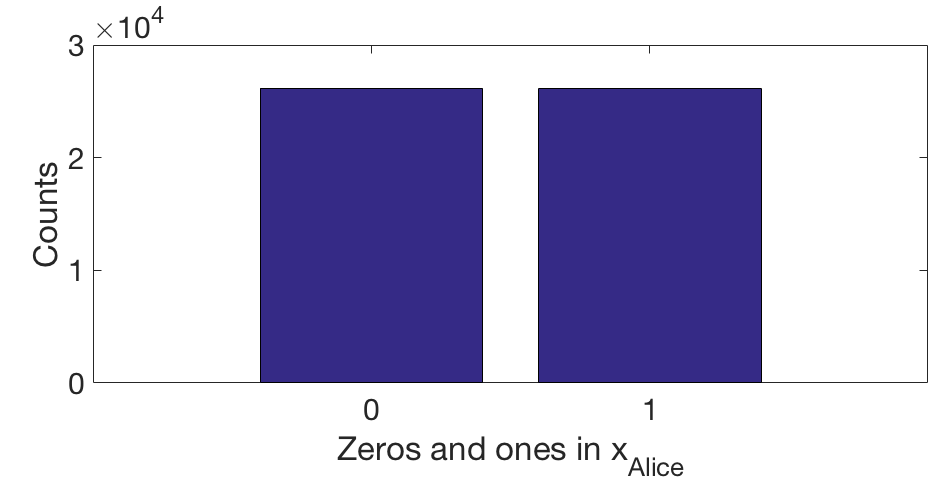}
	\caption{Bit distribution in $x_{Alice}$ based on evaluation dataset collected in Section~\ref{sec:exp}.}
	\label{fig:ones}
	    \vspace{-0.25in}
\end{figure}

The extreme case to consider a binary vector as a non-sparse vector is that it has exactly 50\% ones and 50\% zeros. Figure~\ref{fig:ones} shows that $x_{Alice}$ usually contains approximately 50\% ones and 50\% zeros after quantization, which implies that $x_{Alice}$ is \textit{not sparse}. It is extremely hard for a potential attacker to exploit Vulnerability 1. On the other hand, the success rate to exploit Vulnerability 2 depends on how Eve can approximate inherent IPI without the contact of user's body. In Section~\ref{sec:model_attack}, we assume Eve can perform three types of attacks.

We now quantitatively demonstrate that these two vulnerabilities can be addressed by properly choosing the value of parameter $M$. Let us define $P = S_{\Delta A, B} * log(N/S_{\Delta A, B})$ and $Q = min(S_{Alice}, S_{\Delta A, E})$, respectively. We have:
\newtheorem{cscondition}{Corollary}
\begin{cscondition}
\label{corollary1} 
	\emph{The CS-based reconciliation method is perfectly
		effective (i.e., Bob can recover $\Delta x_{A, B} $ and $x_{Alice}$ successfully)  
		and secure (i.e., Eve is unable to recover $\Delta x_{A, E}$ and $x_{Alice}$ successfully) if}
	\begin{equation}
	P < M < Q
	\label{eq:corollary}
	\end{equation}
\end{cscondition}

\textbf{Proof:}   If Bob cannot recover $\Delta x_{A, B}$ and $x_{Alice}$ with 
a $\Phi$ that satisfies Condition \textbf{C1} and $M > P$, it is contradictory to Condition \textbf{C2}, which defines 
the \emph{sufficient} condition for successful CS decoding with $\ell_1$ minimization.
If Eve can recover $\Delta x_{A, E}$ and $x_{Alice}$ with $M < Q$, it is contradictory to Condition \textbf{C3}, which defines the
\emph{necessary} condition for successful CS decoding with $\ell_1$ minimization
for both \textbf{Vulnerabilities} 1 and 2 because $Q$ is the minimum of 
$S_{Alice}$, $S_{\Delta A, E}$, which are the necessary conditions for 
successful $x_{Alice}$ recoveries in \textbf{Vulnerabilities} 1 and 2 respectively.

\begin{figure}[ht]
	\centering
	\includegraphics[width = 8cm]{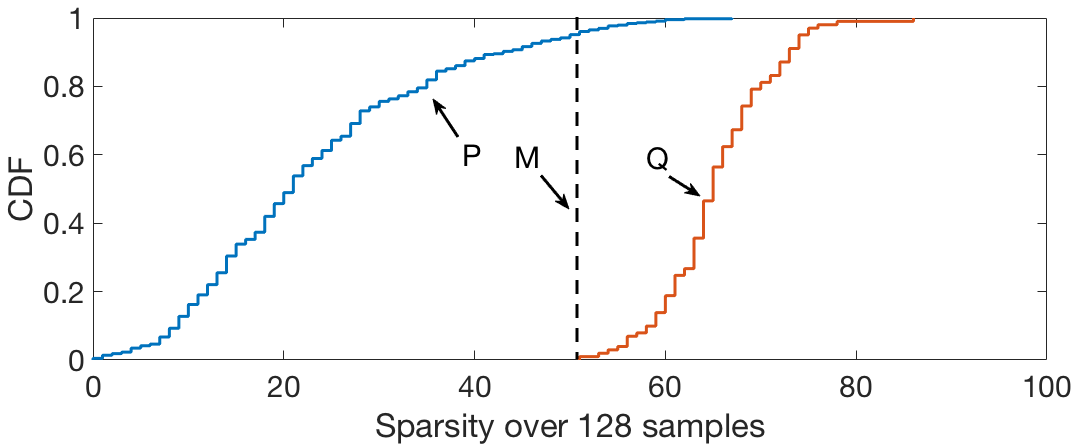}
	\caption{Sparsity when $N = 128$.}
	\label{fig:spar}
	    \vspace{-0.25in}
\end{figure}

The design for an effective and secure CS-based reconciliation becomes a problem to find a suitable $M$. The upper bound $Q$ is the secure threshold and the lower bound $P$ is the effective threshold. These two thresholds are determined by the population of mismatches among legitimate devices and attackers. Assume that H2B aims to generate a $128$-bit key, i.e., $N = 128$. 
Figure~\ref{fig:spar} shows the distribution of $P$ and $Q$ in our dataset introduced in Section~\ref{sec:exp}.
While we cannot find an optimal value of $M$ that satisfies Corollary~\ref{corollary1}, we deliberatively selected a small $M$ value of 50 so that $M < Q$ 
to cover Vulnerabilities 1 and 2. However, there exists a small chance ($4.8\%$) that $M \leq P$, implying that there is $4.8\%$ possibility that the proposed CS-based reconciliation fails to correct all mismatches between $x_{Alice}$ and $x_{Bob}$. Note that we just consider two possible vulnerabilities and three types of attacks. Corollary 1 and the condition $M < Q$ aim to protect two defined vulnerabilities only. 

As the attacker Eve has the full control of the public communication channel, she may modify $y_{Alice}$ in the reconciliation step. To maintain message integrity, a message authentication code (MAC) method is implemented to verify the message ~\cite{bellare1996keying}. Here, H2B codes $y_{Alice}$ to $L_{Alice} = \{y_{Alice}, MAC(x_{Alice}, y_{Alice})\}$, by treating $x_{Alice}$ as the shared secret. On receiving $L_{Alice}$, Bob can obtain $x_{Alice}$ as discussed above and also protect the integrity and the authenticity of the message via the MAC method. If the message is modified, the derived $x^\prime_{Bob}$ will be different from $x_{Alice}$ so that Bob cannot produce the correct $L_{Alice}$. The message is thereby discarded. The MAC method also allows Bob to examine the reconciliation results. As shown in Figure~\ref{fig:ptcl_rec}, Bob will notify Alice for the final result to determine whether the key generation is successful. Finally, we apply SHA2-256 hashing algorithm on the successfully generated key to further amplify privacy.
\section{Evaluation}
\label{sec:evaluation}
\subsection{Prototype Device}
\label{sec:hardware}
\begin{figure}[h!]
	\includegraphics[width = 8cm]{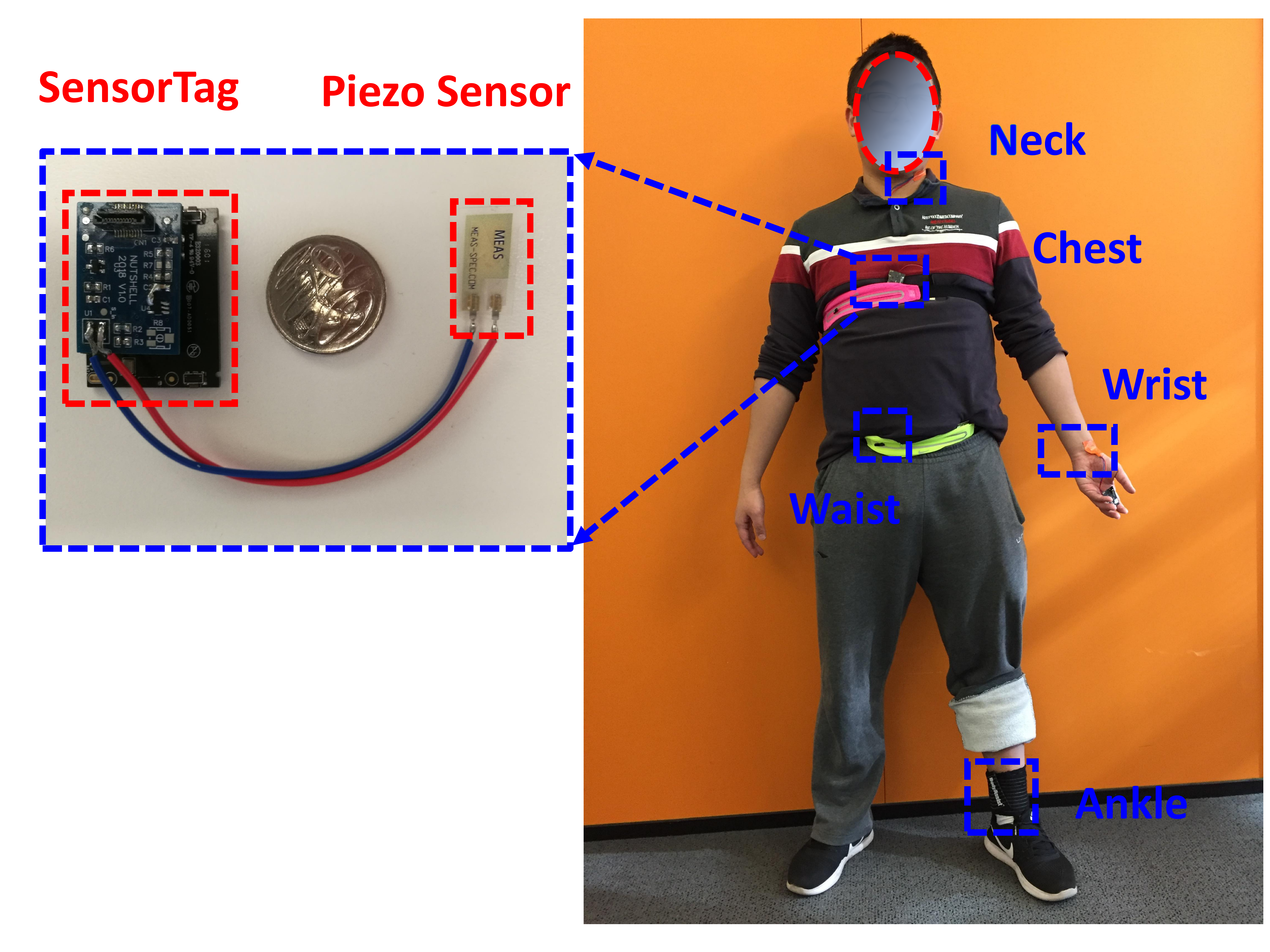}
	\caption{Prototype and data collection.}
	\label{fig:datacollection}
\end{figure}

Among known piezoelectric sensors,  PVDF piezopolymer is usually used in low-frequency applications~\cite{wang2011flexible,chiu2013development}. Therefore, we build the prototype of H2B using off-the-shelf PVDF piezopolymer sensors ($25 \times 13 \times 1mm^3$, US\$5.6, see Figure~\ref{fig:datacollection}), produced by TE Connectivity\footnote{Piezo sensor LDT0-028K: http://www.te.com/global-en/product-CAT-PFS0006.html}. The sensor features $180$ Hz resonant frequency and $50$ mV/g baseline sensitivity. The data logger used is the SensorTag produced by Texas Instruments\footnote{SensorTag: http://www.ti.com/ww/en/wireless\_connectivity/sensortag2015 /index.html}. Piezo sensors are sampled by the on-board $12$-bit Analog-to-Digital (ADC), and all data samples are stored on the onboard flash memory. 

\vspace{-0.2in}
\subsection{Data Collection}
\label{sec:exp}
We collect a dataset using the designed prototype device\footnote{Ethical approval has been granted by the corresponding organization (Approval Number HC17008)}.  The dataset
consists of 23 healthy subjects (7 females and 16 males) aged from 24 to
40 with different skin tones. During data collection, we attach 5 prototype devices on the participant's neck, chest, wrist, waist and ankle respectively as shown in Figure~\ref{fig:datacollection}. The participants are asked to stand static, 
though we find that other still activities such as sitting and lying does not affect the performance. However, H2B does not apply to activities involving large body movements such as walking and running, and We will discuss this limitation in Section~\ref{sec:discussion} later. Each experiment lasts for 5 minutes and the sampling rate is 400Hz.

\subsection{Impact of Sampling Rate}
\label{sec:ev_fs}
We find that the sampling rate will affect how much entropy that can be extracted from measured IPI values. This is because the sampling rate determines the precision of measurement: the higher the sampling rate is, the more accurate the measured IPI is, so is noise. Therefore, the measurements will have more entropy with the increase of sampling rates. The measurements contain two parts: useful IPI (used to generate keys) and noise. As the entropy is a physical feature of a source, it has an upper limit. Although the entropy of measurements increases as sampling rate increases, the entropy of the generated key will not increase after it reaches a maximum value. In other words, there exist a threshold for sampling rate over which the entropy of the generated keys will not improve. To investigate this threshold, we downsample the dataset from 400Hz to 40Hz 
to find out how much entropy can be extracted for different sampling rate.

\begin{figure}[h!]
	\centering
	\includegraphics[width = 8cm]{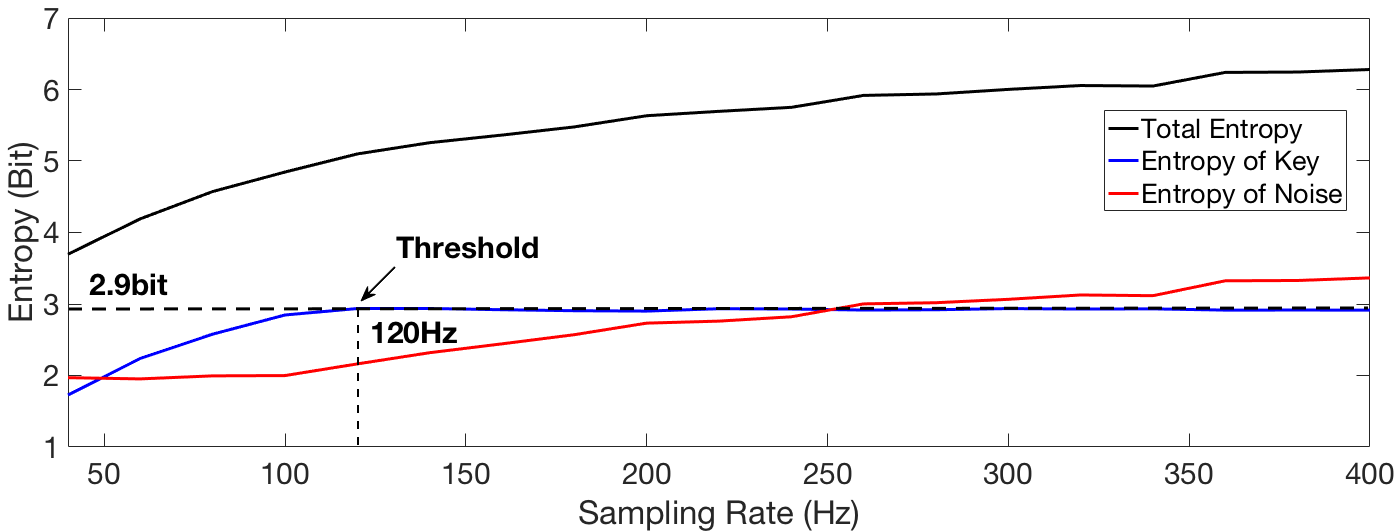}
	\caption{Impact of sampling rate on entropy.}
	\label{fig:en_fs}
	    \vspace{-0.15in}
\end{figure}
Figure~\ref{fig:en_fs} illustrates the impact of sampling rate on entropy. We can see  that the entropy of measurements is positively related to the sampling rate as expected.  After we separate the entropy of generated keys and that of random noise, we find that the entropy of generated keys reaches an upper limit (2.9 bits) and do not increase further. The results suggest that even if we increase sampling rates, the entropy of the generated keys will not increase. Therefore, we set sampling rate to 120Hz in H2B to save energy for resource-constrained wearable devices. In subsequential experiments and evaluation sections, we downsample the raw dataset from 400Hz to 120Hz to examine the performance at the desired sampling rate. 

\subsection{CS-based Reconciliation v.s. ECC}
\label{sec:performance}
We benchmark the performance of the proposed CS-based reconciliation method against the state-of-the-art Error Correction Code (ECC)-based reconciliation~\cite{mathur2011proximate, yang2016secret, xu2016walkie} using the collected dataset. Among various ECC methods, we select (15,3) Reed-Solomon (RS) code as it shows better performance in key reconciliation process than others as described in~\cite{xu2016walkie}. 

The major performance metric for symmetric key generation system is the probability that two keys generated by Alice and Bob can completely agree with each other. The probability of $100\%$ matching, or noted as success rate, is an important metric because only 1 bit mismatching can cause protocol failure. As a result, the key generation process has to re-start which results in prolonged delays, user dissatisfaction and extra device energy consumption.

As most current cryptographic systems require key length of 128 bits, we compare the success rate of CS-based reconciliation and ECC-based reconciliation to generate 128-bit keys.
Evaluation results shows that the success rate of CS-based reconciliation approach is $95.6\%$ while RS(15,3) achieves $34.2\%$ success rate only. The poor performance of ECC-based reconciliation approach on this dataset is due to the high mismatched rates after quantization process that exceed the correction ability of ECC. 

\subsection{Impact of Different Body Locations}
\label{sec:body_location}
\begin{figure}[h!]
	\centering
	\includegraphics[width = 8cm]{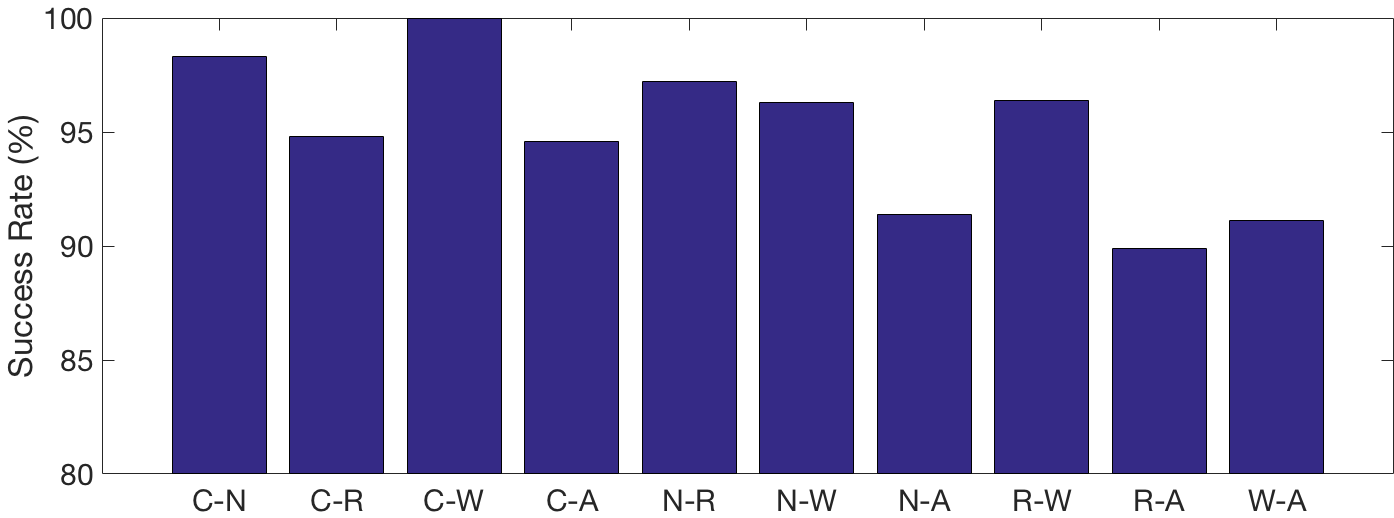}
	\caption{The success rate of different pairs of body locations: C-chest, N-neck, R-wrist, W-waist, A-ankle.}
	\label{fig:location_barplt}
    \vspace{-0.2in}
\end{figure}

\begin{figure}[h!]
	\centering
	\includegraphics[width = 8cm]{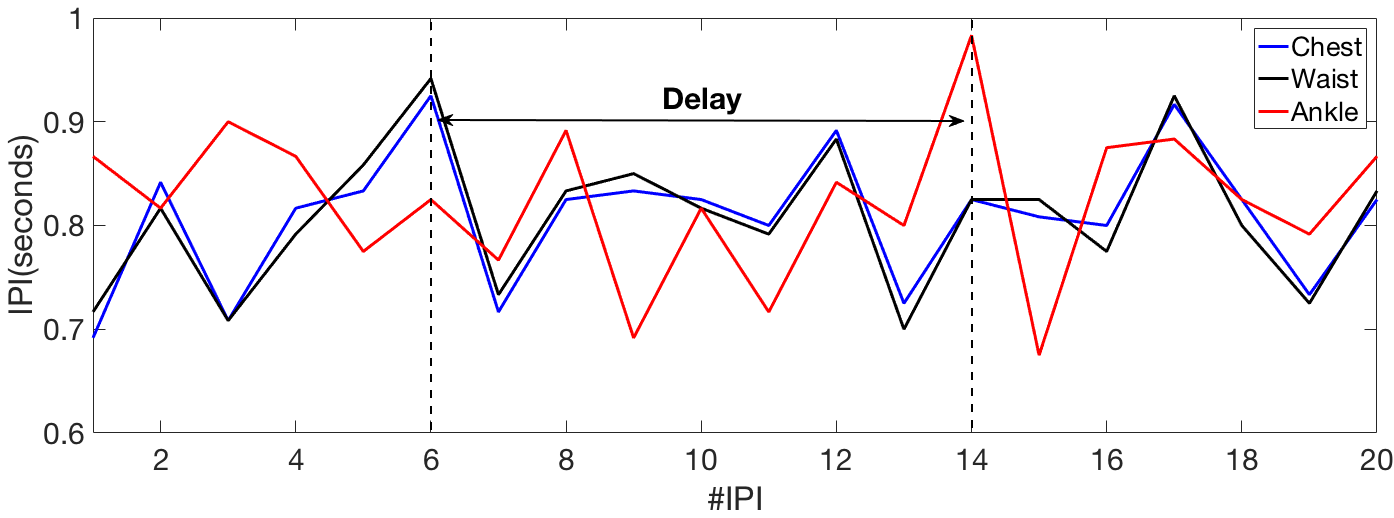}
	\caption{Time delay in IPI measured on ankle from that measured on chest or waist.}
	\label{fig:delay}
    \vspace{-0.25in}
\end{figure}

We examine the success rates of H2B for the devices on five different body locations: chest, neck, wrist, waist, and ankle, and Figure~\ref{fig:location_barplt} shows the success rates on all pairing combinations. The highest success rate is $100\%$ when pairing a device on chest with another on waist because the chest and the waist are close to each other, and both piezo sensors on these two locations  measure SCG signal. We also find that all pairing attempts with the device on the ankle result in success rates that are lower than $95\%$
(though larger than 90\%). This is because ankle is the farthest location from the heart and there is a tiny delay when heartbeat signal is propagated from the chest (heart) to the ankle as shown in Figure~\ref{fig:delay}. 
\subsection{Randomness of Keys}
To validate the randomness of H2B keys, we apply the NIST statistical test~\cite{rukhin2001statistical} to the generated keys after quantization. P-values in Table 5 represent the probability that the dataset are generated from a random process. If a p-value is less than a threshold (usually 1\% or 5\%), the randomness hypothesis is rejected. Table 5 shows that all NIST tests have p-values larger than 1\%. However, as a truly random source would have p-values from NIST tests uniform on [0,1] as described in~\cite{rukhin2001statistical}, future work could provide more evidence on whether or not H2B keys truly meet the randomness condition. 
\begin{table}[ht]
	\centering
	\caption{NIST Test Results}
	\label{tab:nist}
	\resizebox{5cm}{!}{
		\begin{tabular}{cc}
			\hline
			NIST Test  & p-value \\
			\hline
			Frequency  & 0.095581 \\
			FFT Test &  0.046782 \\
			Longest Run & 0.050684 \\
			Linear Complexity  & 0.985610\\
			Block Frequency & 0.208905  \\
			Cumulative Sums & 0.098443  \\
			Approximate Entropy & 1   \\
			Non overlapping Template & 0.018382  \\
			\hline
	\end{tabular}}
    \vspace{-0.2in}
\end{table}

\subsection{Resiliency against Attacks}
\label{sec:ev_security}
As discussed in Section~\ref{sec:model_attack}, we model three types of attacks, by which Eve tries to derive a key. Here, we analyze the security of H2B against these attacks. The metric used in this section is the average bit agreement rate, which is different from success rate in previous evaluation. Bit agreement rate is the portion of matching bits over all bits in a key. A successful key generation process needs to produce a symmetric key with $100\%$ agreement rate.

\subsubsection{Passive Attack}
In an passive attack, Eve attempts to pair her device with the legitimate devices on a user by using her own heartbeat data. We simulate this scenario by pairing devices from the same body location but different users. For example, we pair the device on chest of subject 1 to the devices on the chest of other subjects to see if they can generate the same key. The results of passive attack is shown in Figure~\ref{fig:pa}. We can see that passive attack can at most achieve $70\%$ agreement rate. Therefore, different users produce distinguishable heartbeat signals and H2B utilizes this characteristic to generate different keys.  

\subsubsection{Active Presentation Attack}
\begin{figure}[ht]
	\includegraphics[width = 8cm]{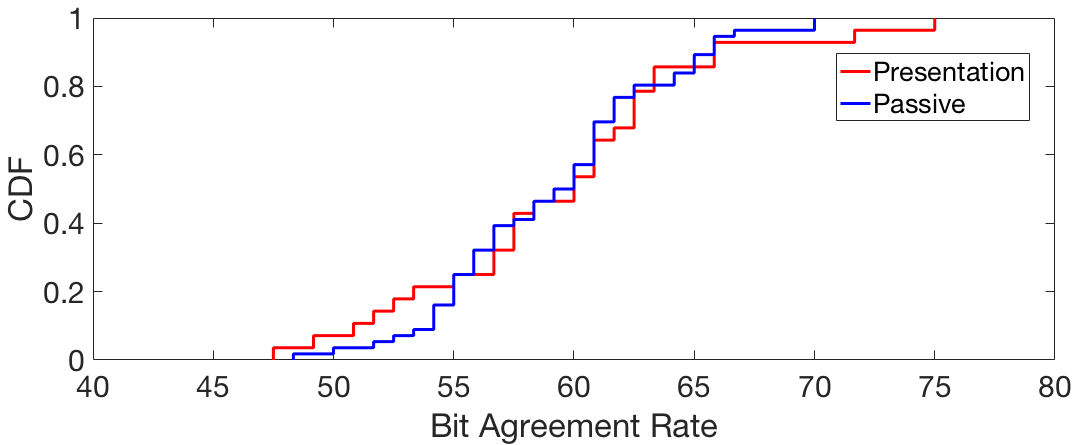}
	\caption{Passive and presentation attacks.}
	\label{fig:pa}
	    \vspace{-0.25in}
\end{figure}

In presentation attacks, we assume Eve can access the historical IPI data of a user and use these data to pair with the legitimate devices of the user in current time. To simulate this scenario, we segment the data of each user into two equal parts.
The first half part is used as historical data to pair the second half part of the same user. The results in Figure~\ref{fig:pa} show that presentation attacks can achieve slightly higher agreement rate than passive attacks because the data are form the same user. However, it can at most achieve $75\%$ agreement rate which in turn demonstrates the randomness of IPI over time and hence the security of H2B against this type of attacks.
\subsubsection{Active Video Attack}
\begin{figure*}
	\subfigure[Original frame.]{ 
	\label{fig:v1} 
	\begin{minipage}[b]{0.3\textwidth} 
		\centering 
		\includegraphics[width=3.5cm]{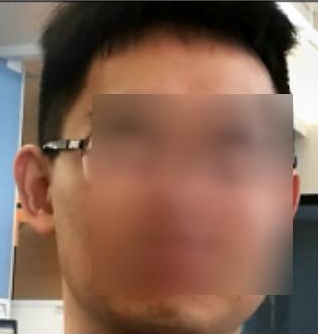}
\end{minipage}}%
	\subfigure[Magnified frame with red box as ROI.]{ 
	\label{fig:v2} 
	\begin{minipage}[b]{0.3\textwidth} 
		\centering 
		\includegraphics[width=3.5cm]{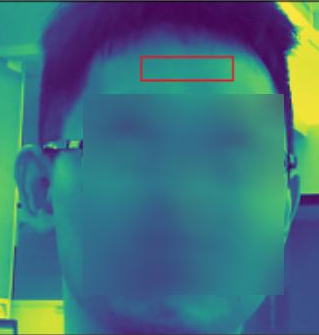}
\end{minipage}}%
	\begin{minipage}[b]{0.4\textwidth} 
	\subfigure[Pixel intensities in ROI.]{ 
	\label{fig:v3} 
		\centering 
		\includegraphics[width=5cm]{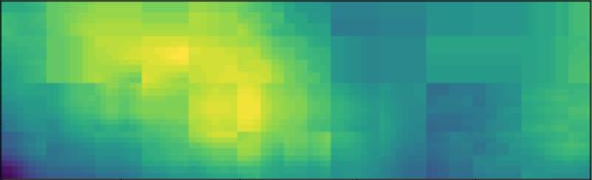}}
	\\
		\subfigure[Intensity curves and extracted peaks.]{ 
		\label{fig:v4} 
		\centering 
		\includegraphics[width=5cm]{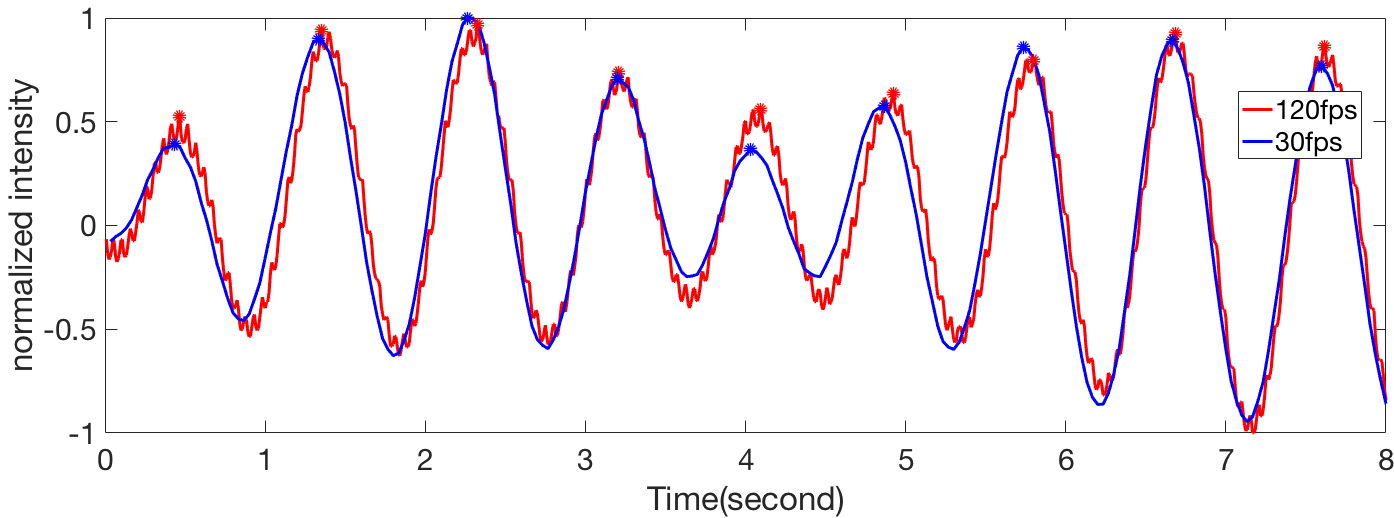}}
\end{minipage}%
	\caption{Active video attack. }
	\label{fig:a_frame}
	    \vspace{-0.22in}
\end{figure*}

Recently Wu et. al. \cite{wu2012eulerian} developed a video magnification procedure that is capable of revealing bio-signals from a video footage of a person. The video is decomposed spatially and temporally to amplify variations occurring within a specified frequency (e.g., heart beating frequency range in our context). The technique has been employed in many applications including adult and infant heart beating tracking\cite{wu2012eulerian, he2014using}. We assume that an attacker Eve can film a video of user's face and apply such technique to extract IPI data. She then use the extracted IPI data to launch an active attack attempt. 

To simulate this attack, we ask a user to directly face a video camera situated at a very close distance to produce
the best IPI results, which is an unfair advantage given to Eve deliberately. A video of the user's face is analyzed using Eulerian Video Magnification~\cite{wu2012eulerian} to extract the heart beating details as shown in Figure~\ref{fig:a_frame}. The video frame rate is set to $30$ and $120$ fps respectively and the region of interest (ROI) is set to part of the user's forehead. Due to the presence of noise in the video, the amplification is implemented by using a narrow band filter $0.5$ - $1.4$ Hz as recommended in~\cite{wu2012eulerian}, and the magnification factor is $60$. 

By tracking the changes in mean intensity of pixels in ROI, Eve can obtain a heartbeat curve as shown in Figure~\ref{fig:v4}. The raw IPIs are extracted using a peak detection algorithm. According to~\cite{balakrishnan2013detecting}, the magnification method for detecting heart rates is not accurate as ECG biosensors. On the other hand, as discussed in Section~\ref{sec:bg_ipi}, the performance of piezo sensor is close to ECG biosensors. Therefore, we can expect that the IPIs estimated from video attacks are different from that measured by piezo sensors. Indeed, the estimation differences of piezo sensor and video attacks can be up to $30\%$ on our dataset.

By using the IPI values estimated from video analysis to pair with those produced by the piezo sensor, the agreement rate is $64.53\%$ for 30fps video, and $66.67\%$ for 120fps video, respectively. This result validates our analysis above, i.e., the IPI values extracted from the video do not have good proximity with those from the piezo sensors. Moreover, to process a 90s video (approximate time of sufficient number of IPI values for H2B to produce 128-bit keys), the processing time is 16 minutes for the 30fps video, and 75 minutes for the 120fps video
respectively with a computer that has 8GB RAM and an Intel i7 processor. Therefore, the valid period of the
secret keys may be set to a short duration (e.g.,  5 minutes) to protect against this types of attacks. 

Finally, as in passive and active presentation attacks, active video attacks have 0 success rate. The analysis indicates H2B is highly attack-proof on three types attacks.

\subsection{Power Consumption}
\label{sec:power}
Power consumption of H2B can result from three parts: data sampling, data processing, and data transmission. Due to recent development in ambient backscatter communication~\cite{liu2013ambient, kellogg2014wi}, the energy consumption of wireless communication is negligible for transmission. Therefore, we focus on evaluation of power consumption in data sampling and processing only. 

\subsubsection{Power Consumption of Data Sampling}
\label{subsubsec:power}
\begin{figure}
	\centering
	\includegraphics[width=6cm]{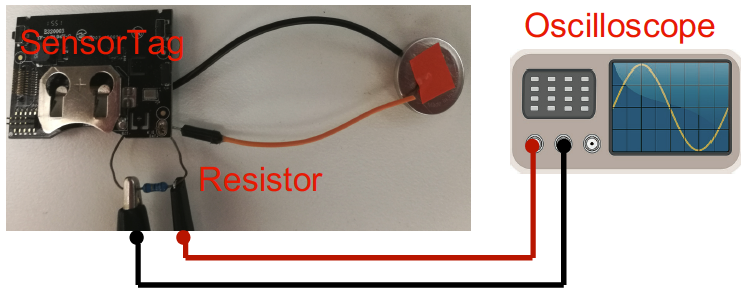}
    \caption{Power consumption of data sampling.}
	\label{fig:measurement_setup}
	    \vspace{-0.2in}
\end{figure}

In our power measurement experiment, we connect the prototype device to a GDS-800 digital oscilloscope~\footnote{GDS-800: http://www.gwinstek.com/en-GB/products/GDS-800} to measure the average power consumption for each sampling event. The measurement setup is shown in Figure~\ref{fig:measurement_setup}. The SensorTag was running with the latest version of Contiki OS, in which the MCU was duty-cycled to save power. Moreover, all unnecessary components, including the ADC, SPI bus, and on-board sensors were powered-off when possible. 

Our measurement shows that the sampling rate of $120$ Hz results in $491.55\mu W$ power consumption in sampling the heartbeat signal. The capacity of a typical battery for SensorTag, such as Panasonic CR2032, is $3$ V, $225$ mAh which is equivalent to $2.43 \times 10^6$ mJ. Therefore, a typical battery can support piezo data continuously sampling at $120$ Hz for $4.96 \times 10^6$ s ($\approx$ 57 days). 

\subsubsection{Time and Energy Consumption of Reconciliation}
The most computational intensive operation in H2B is CS-based reconciliation. However, although $\ell_1$ optimization is known to be computationally intensive, the reconciliation only requires to be executed in \textit{one of the two devices}. It enables H2B to exploit the heterogeneity of wearable IoT architecture. For example, during the key generation process, we can choose the device with more processing power such as smartwatch to run reconciliation process. The less powerful devices such as IMDs needs to transmit messages only. Therefore, we evaluate the power consumption of reconciliation on two modern wearable devices: Samsung Smartwatch and Google Nexus 4. The size of sensing matrix $\Phi$ is empirically chosen to be $50 \times~128$ as discussed in Section~\ref{sec:rec}. The $\ell_1$ optimization used in the system is $\ell_1$-Homotopy~\cite{donoho2008fast}. 
\begin{table}[h]
	\centering
	\caption{Time and energy consumption of reconciliation.}
	\label{table:pc_cs}
	\resizebox{7cm}{!}{
		\begin{tabular}{ccc}
			\toprule
			Device& \textbf{Time(ms)}& \textbf{Energy(mJ)} \\ \hline
			Samsung Smartwatch & 135 & 83.8\\
			Google Nexus 4 & 155 & 143.2\\
			\bottomrule
	\end{tabular}}
	    \vspace{-0.1in}
\end{table}

 Table \ref{table:pc_cs} shows the mean result of the time and energy consumption in one reconciliation process. The results are obtained by running reconciliation 50 times and measured by a widely used open source tool PowerTutor\footnote{http://ziyang.eecs.umich.edu/projects/powertutor/}. Let us take Google Nexus 4 as an example. According to the results, CS-based reconciliation takes $155$ ms and $143.2$ mJ on average. The average key generation rate of H2B is approximately 3 $bit/s$. Therefore, it takes approximately 42.7s to generate a 128-bit key. Consequently, the processing time of $155$ ms for CS-based reconciliation is only a small fraction of the total key generation process. The battery capacity of Google Nexus 4 is $3.8$ V, $2100$ mAh which is equivalent to $2.9 \times 10^4$J. For example, if a user run H2B for 24 hours and generate a new key every 5 minutes, which can be considered as an extreme case, the energy consumed by CS-based reconciliation is $0.09\%$ of the total energy in Google Nexus 4 only.

\section{Discussion}
\label{sec:discussion}
\subsection{Piezo Sensor v.s.  ECG}
In this subsection, we benchmark the piezo sensor against ECG biosensor. For ECG biosensor,  we use Neurosky's TGAM1 ECG biosensor\footnote{http://neurosky.com/biosensors/ecg-sensor/} as the baseline. As shown in Table~\ref{table:piezovsecg}, the advantage of piezo sensor is that it is smaller, more light, and more cost and power efficient compared to ECG sensors. Moreover, piezo sensors do not require skin contact when they measure SCG on the chest or waist while ECG sensors always need skin contact. 
\begin{table}[h!]
	\centering
	\caption{Comparison between piezo and ECG sensor.}
	\label{table:piezovsecg}
	\resizebox{7cm}{!}{
		\begin{tabular}{ccccc}
			\toprule
			&   Size & Weight & Power & Price \\
			&    ($mm^{3}$) & (g) & (mW) &(US\$)\\ \hline
			Piezo Sensor& $25 \times 13 \times 1$ & 0.05& 0.49  & 5.6 \\
			ECG Sensor  &$27.9 \times 15.2 \times 2.5$& 0.13 & 2.5 & 55 \\
			\bottomrule
	\end{tabular}}
	    \vspace{-0.2in}
\end{table}

Apart from piezo and ECG sensors, IPI values can be also measured by PPG sensors and accelerometers. PPG sensors that require the operation of LED that consumes more power. A state-of-the-art PPG sensor AnalogDevices ADPD2212 \footnote{http://www.analog.com/cn/products/adpd2212.html} consumes power of $2.8mW$ in operation. A recent study~\cite{xu2017keh} shows that piezo sensor-based gait recognition system can save by $78.15\%$ compared to accelerometers. Therefore, H2B shows advantages over such sensors in terms of power savings. 

\subsection{Limitations}
In this paper, we show the potential of using wearable piezo sensors to detect heartbeat signal and further utilize IPI information to generate keys. Although piezo sensors are not widely used in common wearable devices at the moment, more and more commercial devices have emerged in the market as discussed in Section~\ref{sec:introduction}. With the advancement of technology especially self-powering techniques, PEH or piezo sensors will be embedded in more future wearable devices. Therefore, we believe H2B makes a humble step towards securing the communication of these wearable devices.

Another limitation is that the key generation rate of H2B is 2.9 bit per IPI. Therefore, it takes approximately 40 seconds to generate a 128-bit key as a typical heart rate is 60 - 100 beats per minute~\cite{obrist2012cardiovascular}. In fact, this is a common problem in IPI-based key generation system because the limited entropy contained in the IPI. For example, the key generation rate of two recent work H2H~\cite{rostami2013heart} and IMDGuard~\cite{xu2011imdguard} is 4 bit per IPI. The lower key generation rate of H2B is because the signal measured by piezo sensors has low SNR.

In the current design, H2B only works when a user is in static activities such as sitting, standing still and lying. This is because piezo sensors are sensitive to all kinds of body motion artifacts. Therefore, the large motions will overwhelm the weak heartbeat signal when the user is moving. To address this problem, more advanced signal processing approaches are required such as source separation technique, which we leave as future work. 

In Section~\ref{subsec:rec}, we defined two possible vulnerabilities and three attack attempts and  found a security boundary $M < Q$ to protect these vulnerabilities. However, there may be other potential vulnerabilities due to the information leakage and there is also a possibility that effective attacks exploiting Vulnerability 2 may be developed in the future. Future work could provide comprehensive study on computational secrecy of CS-based reconciliation protocol theoretically.


\section{Related Work}
\label{sec:related_work}
Biometrics has been well developed in two types of applications in the security of wearable devices, i.e. user authentication and symmetric key generation. Authentication systems~\cite{xu2017keh, wang2018unlock} exploit a unique pattern (distinctiveness) for each individual. The pattern must be \textit{time-invariant} so that the user can be identified by the corresponding feature set. Therefore, authentication systems usually normalize biometrics to remove randomness over time. On the contrary, symmetric key generation systems~\cite{xu2016walkie, schurmann2017bandana} need to exploit both distinctiveness and randomness of biometrics. Randomness over time (forward secrecy) can ensure that an attacker cannot guess the key from past data. Therefore, a symmetric key generation system usually extracts \textit{time-variant} biometric features. Our work is a symmetric key generation system so we highlight this difference.

Biometrics is the most popular trend in key generation for BAN. Among all biometrics, IPI is the most common choice of key material. This trend starts from Poon et. al~\cite{poon2006novel} when they exploited IPIs as the entropy source. The systems exploited IPIs include IMDGuard~\cite{xu2011imdguard}, OPFKA~\cite{hu2013opfka}, ESDS~\cite{zheng2014securing}, Heart2Heart~\cite{rostami2013heart}. The previous works focus on either ECGs or PPG signals to extract IPIs. To the best of our knowledge, using IPI measured by piezo sensor has not been studied. To this end, H2B presents a humble first step 
to develop a cryptographic system based on IPIs extracted from noisy piezo signals. 

\section{Conclusion}
\label{sec:conclusion}
We proposed and implemented a shared secret key generation system, called H2B, for secure BAN communications. H2B allows two wearable devices to dynamically generate the same key on-the-fly based on the IPI measured by inexpensive and energy efficient but noisy piezo sensors. By employing an quantile-based quantization and a CS-based reconciliation, H2B can successfully pair two wearable devices on the same user's body with a probability of $95.6\%$. We also demonstrate that H2B is secured against three types of typical attacks and very power-efficient.   

\begin{acks}
The authors wish to thank Jimmy Chan from WBS Technology for hardware support, Guohao Lan and Dong Ma for experiment assistance, and the anonymous reviewers for providing valuable comments. 
\end{acks}

\bibliographystyle{ACM-Reference-Format}
\bibliography{contents/bib_ipi}

\end{document}